\documentclass[journal,10pt]{IEEEtran}

\usepackage{amsmath}
\usepackage{bbold}
\usepackage{enumerate}
\usepackage{cite}
\usepackage{color,xcolor}
\usepackage{amssymb}
\usepackage{graphicx}
\usepackage{algorithm}
\usepackage{algpseudocode}
\usepackage{bbm}
\usepackage{graphics}
\usepackage{subfigure}
\usepackage{epsfig}
\usepackage{epstopdf}

\newtheorem{theorem}{Theorem}{}
\newtheorem{proposition}{Proposition}{}
\newtheorem{lemma}{Lemma}{}
\newtheorem{assumption}{Assumption}
\newtheorem{corollary}{Corollary}{}
\newtheorem{remark}{Remark}{}
\newenvironment{proof}{\hspace{0ex}\textsc{Proof}.\hspace{1ex}}{\hfill$\Box$\newline}

\begin{document}

\title{Linear Quadratic Regulator Design for Multi-input Systems with A Distributed Cooperative Strategy}

\author{Peihu Duan,~Zhisheng~Duan,~Lidong~He, ~and
        ~Ling~Shi

\thanks{P. Duan and L. Shi are with the Department of Electronic and Computer Engineering, the Hong Kong University of Science and Technology, Clear Water Bay, Kowloon, Hong Kong, China.  E-mails: duanpeihu.work@gmail.com (P. Duan), eesling@ust.hk (L. Shi).  }
\thanks{ Z. Duan is with the State Key Laboratory for Turbulence and Complex Systems, Department of Mechanics and Engineering Science, College of Engineering, Peking University, Beijing, 100871, China. E-mail: duanzs@pku.edu.cn.}
\thanks{L. He is with the School of Automation, Nanjing University of Science and Technology, Nanjing 210094, China. E-mail: lidonghe@njust.edu.cn.}
}

\maketitle

\begin{abstract}
In this paper, a cooperative Linear Quadratic Regulator (LQR) problem is investigated for multi-input systems, where each input is generated by an agent in a network. The input matrices are different and locally possessed by the corresponding agents respectively, which can be regarded as different ways for agents to control the multi-input system. By embedding a fully distributed information fusion strategy, a novel cooperative LQR-based controller is proposed. Each agent only needs to communicate with its neighbors, rather than sharing information globally in a network. Moreover, only the joint controllability is required, which allows the multi-input system to be uncontrollable for every single agent or even all its neighbors. In particular, only one-time information exchange is necessary at every control step, which significantly reduces the communication consumption. It is proved that the boundedness (convergence) of the controller gains is guaranteed for time-varying (time-invariant) systems. Furthermore, the control performance of the entire system is ensured. Generally, the proposed controller achieves a better trade-off between the control performance and the communication overhead, compared with the existing centralized/decentralized/consensus-based LQR controllers. Finally, the effectiveness of the theoretical results is illustrated by several comparative numerical examples.
\end{abstract}


\section{Introduction} \label{sec1}
The past several decades have witnessed a considerable research boom in networked systems for complex tasks, such as the collaborative transportation by multiple robots \cite{michael2011cooperative,marino2017distributed}, the cooperative detection by networked sensors \cite{bliss2014cooperative,yang2014stochastic}, and the surgical operation by several manipulators \cite{lanfranco2004robotic}. Compared with a single system, networked systems possess great superiority in rich functionality and system robustness.

To accomplish tasks effectively and efficiently, one significant issue is to design the cooperative strategy, which has been investigated by a large number of existing works, such as consensus control \cite{olfati2007consensus}, containment control \cite{meng2010distributed}, formation control \cite{oh2015survey}, etc. These consensus-based strategies mainly aim at a global objective that each system in a network converges to a consistent view of their states.
However, for some tasks where networked systems cooperatively assist a plant in tracking a given trajectory, the state of each system may need to be inconsistent. For example, when multiple robots rotate a large object collaboratively, the forces applied to the object and the positions of different robots are different. In these cases, one critical issue is to design the required force that needs to be exerted on the object by each system. An effective way to model this issue is to regard the plant as a dynamical multi-input system \cite{lavaei2009overlapping}, where each input is generated by an agent in a network. These networked agents can interact with neighbors in a communication graph.

To deal with the above control problem of multi-input systems, a large number of methods have been developed, divided into centralized, decentralized and distributed ones. By collecting full information, centralized methods can develop the optimal control input for each agent \cite[Chapter 4]{bertsekas1995dynamic}. However, they pay a great price for communication and computation  when the number of inputs increases.
As a remedy, decentralized methods have been designed in \cite{wang1973stabilization,corfmat1976decentralized,1429376,blanchini2014network}.  Earlier, a class of local control laws, depending on partial system information, were proposed for stabilizing multi-variable systems in \cite{wang1973stabilization,corfmat1976decentralized}. Later, these results were extended to cases with rate constraints on collecting local information \cite{1429376}. Recently, in \cite{blanchini2014network},  structured linear matrix inequalities were adopted to obtain solutions to the control problem of multi-input systems. It should be noted that, although the above decentralized methods can be performed based only on local information, the controller gains have to be designed by a central unit with global information. To locally design the controllers, distributed methods have been proposed \cite{smith2007closed,sturz2020distributed}. For example, cooperative control of vehicles was realized by using a distributed designing method in \cite{smith2007closed}. However, input matrices for all vehicles must be identical. Besides, the methods in \cite{wang1973stabilization,corfmat1976decentralized,1429376,blanchini2014network,smith2007closed,sturz2020distributed} are restricted to stabilizing multi-input systems, and more complex control performances are unclear.

Since an LQR performance index is effective to evaluate the control performance and resource, various LQR controllers have been developed for dynamical systems with multiple inputs   \cite{Speyer1979,shoarinejad1999two,6600886,borrelli2008distributed,vlahakis2018distributed,li2015fully,wang2019lq}. As a fundamental result, a class of decentralized linear quadratic Gaussian control laws were designed in \cite{Speyer1979}, whereas local data had to be transmitted from each agent to every other agent. Later, in \cite{shoarinejad1999two}, the stability analysis of a similar problem was intensively studied under two-input cases with noisy communication.  In \cite{6600886}, a near optimal LQR performance was achieved in a decentralized setting.
However, all information of the multi-input system, such as all input matrices, has to be shared for every agent to construct its controller in \cite{Speyer1979,shoarinejad1999two,6600886}. Hence, these methods are more like centralized ones. Recently, distributed LQR control problems, where performance indexes evaluated the behaviors of networked identical and nonidentical systems, were investigated in \cite{borrelli2008distributed} and \cite{vlahakis2018distributed}, respectively. Besides, considering a terminal constraint of consensus, an LQR-based control performance was ensured for multi-agent systems in \cite{li2015fully,wang2019lq}. It should be noted that networked systems considered in \cite{borrelli2008distributed,vlahakis2018distributed,li2015fully,wang2019lq} have to be decoupled into separate subsystems. Specifically, state matrices of these large-scale systems have to be diagonal so that each agent has its own independent system dynamics. However, state matrices of systems with networked inputs usually are in arbitrary forms. Meanwhile, each subsystem is assumed to be completely controllable in these works. Hence, the methods in \cite{borrelli2008distributed,vlahakis2018distributed,li2015fully,wang2019lq} cannot be directly applied for cases considered in this paper. Lately, by introducing an average consensus information fusion strategy, Talebi et al. \cite{talebi2019distributed} designed a distributed control law for multi-input systems with heterogeneous input matrices under the joint controllability. Nonetheless, to deal with local uncontrollability, each agent needs to exchange information infinite times at every step, resulting in a very high communication and computation cost.
Therefore, only based on the joint controllability and limited computation and communication resources, a distributed controller with a global LQR performance index for multi-input systems is still lacking. The technical challenges lie  in two factors: 1) how to establish the relation between the global index and the local control strategy of each agent; 2) how to achieve a better trade-off between limited resources and the control performance.

Motivated by above observations, this paper investigates the distributed LQR-based controller design problem for multi-input systems, where each input is determined by an agent in a network. Since different agents may adopt different manners to exert their inputs, input matrices are considered to be heterogeneous and time-varying. In this sense, these input matrices are regarded as local information, owned by corresponding agents. By embedding a fully distributed information fusion strategy, a new cooperative controller is proposed, where each agent only needs to interact with its neighbors, rather than sharing information globally  over a network. Compared with the literature, this paper possesses the following contributions:
\begin{enumerate}
    \item Each control input of the multi-input system is generated by an agent, based only on its own and neighbors' information. Compared with the centralized methods or the decentralized ones with global information in \cite{Speyer1979,bertsekas1995dynamic,shoarinejad1999two,6600886}, the designing and performing processes of the controller proposed in this paper consume much less  communication resources.
           \\
    \item An LQR performance index of the entire system can be ensured under a milder joint controllability, rather than the complete controllability of every subsystem assumed in \cite{borrelli2008distributed,li2015fully,vlahakis2018distributed,wang2019lq}. This relaxation greatly increases the applicability of the proposed controller.
           \\
    \item Compared with  infinite information exchanges at every step  in \cite{talebi2019distributed} by using the average consensus fusion, only one-time transmission is needed at every step in this paper. Hence, a better trade-off between the control performance and the  communication cost is  achieved by the proposed controller.
  \end{enumerate}

The structure of this paper is given as follows. In Section \ref{sec2}, some preliminaries and the problem statement are provided. In Section \ref{sec3}, an LQR-based controller with a distributed fusion strategy are designed, and a detailed performance analysis  is given.  In Section \ref{sec4}, two simulation examples are presented  to illustrate the effectiveness of the theoretical results. In Section \ref{sec5}, a conclusion is drawn.

\textit{Notations}: Let $ I_n \in \mathbb{R}^{n \times n} $ denote an identity matrix, $\mathbb{1}_n \in \mathbb{R}^{n \times 1}$ denote a vector with all elements being $1$, and $\mathbb{0}_{ n} \in \mathbb{R}^{1 \times n }$ denote a vector with all elements being $0$, respectively. For any matrix $S$ of appropriate dimensions, let $S^{-1}$, $S^{T}$ and $\|S\|_2$ represent its inverse, transpose and 2-norm, respectively. For any positive definite matrix $P$, let $\lambda_{min} (P)$ and $\lambda_{max}(P)$ denote the smallest and largest eigenvalues of $P$, respectively. For two square matrices $A$ and $B$ of appropriate dimensions, $A \ge B$ ($A \leq B$) means that $A - B$ is a positive (negative)  semi-definite matrix. For two positive scalars $\rho_1$ and $\rho_2$, and a positive definite matrix $P$,  $\rho_1 I \leq P \leq \rho_2 I$ indicates that $ \lambda_{min} (P) \ge  \rho_1  $ and $  \lambda_{max} (P) \leq \rho_2  $. For two matrices $S$ and $P$, let $(S)^TP(*) $ denote $S^TPS$.

\section{Preliminaries and Problem Formulation}\label{sec2}

\subsection{Preliminaries}\label{sec2.1}
Let $\mathcal{G}=(\mathcal{N}, \ \mathcal{E})$ denote a communication graph, where $\mathcal{N}=[1,\ \ldots, \ N]$ is the node (agent) set and $\mathcal{E} \subseteq \mathcal{N} \times \mathcal{N} $ is the interaction edge set. In a directed communication graph, an edge $(i, \ j)$ represents that node $i$ can receive information from node $j$, but not necessarily vice versa.
Then, let $\mathcal{A}=[a_{ij}]_{N \times N}$ denote the adjacency matrix, where if $(i, \ j) \in \mathcal{E}$, $a_{ij}=1$; otherwise $a_{ij}=0$. Particularly, since node $i$ can naturally obtain its own information, $a_{ii}=1$, $\forall i \in \mathcal{N}$. Let $\mathcal{N}_i $  denote the in-neighborhood set of node $i$, i.e., $\mathcal{N}_i  = \{j   | (i, \ j) \in \mathcal{E}, \ j \in \mathcal{N}\}$. The in-degree and out-degree of node $i$ are denoted by $d^{in}_{i}$ and $d^{out}_{i}$, respectively. For more knowledge on graph theory, please refer to \cite{bollobas2013modern}.

\begin{assumption} \label{asm:Communication} [Communication topology]
The communication graph $\mathcal{G}$ is directed and strongly connected.
\end{assumption}

\begin{lemma} \label{lemma0} \cite{horn2012matrix}
If Assumption \ref{asm:Communication} holds, $a_{ij,l}>0$, $\forall l \ge N-1,$ $\forall i,j \in \mathcal{N},$ where $[a_{ij,l}]_{N \times N} \triangleq \mathcal{A}^l=[a_{ij}]_{N \times N}^l $.
\end{lemma}

\vspace{6pt}

\begin{proof}
First, according to \cite[Theorem 3.2.1]{brualdi1991combinatorial}, the associated adjacency matrix  $\mathcal{A}$ is irreducible when the communication graph is strongly connected. Then, since $a_{ij} \ge 0$, $\forall i,j \in \mathcal{N}$, $\mathcal{A}$ is nonnegative. Moreover, all the main diagonal entries of $\mathcal{A}$ are positive, i.e., $a_{ii}=1 > 0$, $\forall i \in \mathcal{N}$. Hence, according to \cite[Lemma 8.5.5]{horn2012matrix}, all entries of $ \mathcal{A}^{l}$, $\forall l \ge N-1$, are positive. Thus, the proof of Lemma  \ref{lemma0} is complete.
\end{proof}


\vspace{6pt}

\begin{lemma} \label{lemma1} \cite{zhou1996robust}
For any matrix $P \in \mathbb{R}^{m \times m}$ and vectors $\alpha_{i} \in \mathbb{R}^{m}$, $i=1$, $\ldots$, $N$, the following inequality always holds
\begin{align}
  \bigg ( \sum_{i=1}^{N} \alpha_{i}^T \bigg ) P  \bigg ( \sum_{i=1}^{N} \alpha_{i} \bigg ) \leq  \sum_{i=1}^{N} N \alpha_{i}^T P \alpha_{i}. \notag
\end{align}
\end{lemma}

\subsection{Problem Statement}\label{sec2.2}
Consider a class of multi-input time-varying linear systems, whose dynamics are described by
\begin{align} \label{system_model1}
x_{k+1}= A_{k} x_k + \sum_{i=1}^{N} B_{k,i} u_{k,i},
\end{align}
where $x_k \in \mathbb{R}^n$ is the system state at step $k$, $u_{k,i} \in \mathbb{R}^{m_i}$ is the input governed by agent $i$ at step $k$, $A_k \in \mathbb{R}^{n \times n} $ is the system state matrix, and $B_{k,i} \in \mathbb{R}^{n \times {m_i}} $ is the system input matrix. Here, $B_{k,i}$ is time-varying and local information owned by agent $i$. It is worth mentioning that agents can share information with neighbors in a communication graph $\mathcal{G}=(\mathcal{N}, \ \mathcal{E})$. The above model can characterize cases where $N$ agents drive a plant cooperatively. In the following, two examples are provided to illustrate this formation.

\vspace{6pt}

\begin{figure}[!htb]
\center
\subfigure{{\includegraphics[scale=1.3]{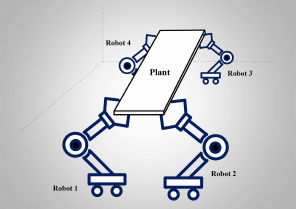}}}
\caption{Four robots perform a carrying task.} \label{f:model_example}
\end{figure}

\vspace{6pt}

\noindent
\textbf{Example 1:} Fig.~\ref{f:model_example} shows that four robots carry a huge object cooperatively, where each robot utilizes its manipulator hand to exert a force on the object. The input force  $u_{k,i}$ is applied by robot $i$ at its own acting points and attitudes, further modeled by $B_{k,i} u_{k,i}$. Since the input matrix $B_{k,i}$ is mainly detected or measured by robot $i$, it is regarded as local information. Thus, by considering the position and velocity of the object as the system state, its evolution can be described by (\ref{system_model1}).

\vspace{6pt}
\noindent
\textbf{Example 2:} For a multi-input system in the augmented form of $ x_{k+1}=A_{k} x_k + B_{k} u_{k} $, where $B_{k} = [B_{k,1}$, $\ldots$, $B_{k,N}] \in \mathbb{R}^{n \times {m}} $ and $u_{k}  = [u_{k,1}^T$, $\ldots$, $u_{k,N}^T]^T \in \mathbb{R}^{m} $ with $m=\sum_{i=1}^{N} m_i$. For some large-scale or higher-order complex systems, such as multiple unmanned aerial vehicles (UAVs), inputs $u_{k,i}$ are usually designed by decoupling the entire system into several subsystems. For each subsystem, the controller $u_{k,i}$ needs to be developed based on its local information.  In such cases, the system dynamics can be modeled by (\ref{system_model1}).

\vspace{6pt}

\begin{assumption} \label{asm:Joint} [Joint Controllability]
$(A_{k}$, $B_{k})$ is uniformly controllable with $B_{k}=[B_{k,1}$, $\ldots$, $B_{k,N}]$, i.e., there exists a positive scalar $\eta$ and a positive integer $L$ such that
\begin{align}
 \eta I_n \leq \sum_{h=0}^{L-1}   \Phi_{k+h-1,k} B_{k+h} R_{k+h}^{-1} B_{k+h}^{T}  \Phi_{k+h-1,k}^{T} ,     \notag
\end{align}
where
\begin{align}
& \Phi_{k+h-1,k}=
  \begin{cases}
     A_{k+h-1} \times \cdots \times A_{k},     & \text{if $ h \ge 1$ }, \\
    I, & \text{if $ h = 0$ },  \\
  \end{cases} \notag
\end{align}
and $R_{k+h}$ is a positive definite matrix to be given in (\ref{JM_R}).
\end{assumption}

\vspace{6pt}

\begin{remark}
Assumption \ref{asm:Joint} holds for all time-invariant systems with $(A$, $[B_{1}$, $\ldots$, $B_{N}])$ being controllable since the controllability Gramian is positive definite \cite{chen1999linear}. This assumption allows the dynamical system to be uncontrollable not only for any agent, but even for the neighborhood of each agent. Specifically, $(A$, $[B_{j_{1}}$, $\ldots$, $B_{j_{d_i}}])$, $j_{1}$, $\ldots$, $j_{d_i} \in \mathcal{N}_i$, can be uncontrollable for all $i \in \mathcal{N}$.  In fact, the joint controllability is almost the mildest condition to guarantee the stability of the multi-input system (\ref{system_model1}), compared with the existing works \cite{borrelli2008distributed,li2015fully,vlahakis2018distributed,wang2019lq}. If even such a mild condition is not satisfied, it is practically impossible to obtain results on the stability of the distributed control for multi-input systems.
\end{remark}

\vspace{6pt}

In the existing works studying system (\ref{system_model1}), one critical issue is the LQR design problem, i.e., how to develop an optimal controller balancing the control resource and the control performance. This problem can be described as
\begin{align} \label{equ:Optp1}
  &  \quad  \min_{\stackrel{u_{k,i}, i \in \mathcal{N}} {k=0, \ldots, M-1 }}  J_{M} ,   \\
 s.t. \quad  & x_{k+1}  = A_{k} x_k + \sum_{i=1}^{N} B_{k,i} u_{k,i}, \notag
\end{align}
where $J_{M} $ is a cost function defined as follows:
\begin{align} \label{JM_R}
& J_{M}
=   x_{M}^T Q_{M} x_{M} + \sum_{k=0}^{M-1} \big \{x_{k}^T Q_{k} x_{k} + u_{k}^T R_{k} u_{k} \big \},
\end{align}
in which  $u_{k}=[u_{k,1}^T$, $\ldots$, $u_{k,N}^T]^T$, $R_{k}=\text{diag}[R_{k,1}$, $\ldots$, $R_{k,N}]$. Here, $Q_{k} \in \mathbb{R}^{n \times {n}} $ and $R_{k,i} \in \mathbb{R}^{m_i \times {m_i}} $, $i \in \mathcal{N}$, are positive definite matrices, characterizing the cost weights of the system state and the input of each agent, respectively.  In the following, three typical frameworks for solving the above optimization problem in the literature are listed.

 \noindent
\textbf{Framework 1:} Usually, the inputs $u_{k,i}$, $i \in \mathcal{N}$, are designed in a centralized manner by a central processor. For example, from \cite[Chapter 4]{bertsekas1995dynamic}, the optimal solution to Problem (\ref{equ:Optp1}) is given by
\begin{align} \label{equ:centralc}
[u_{k,1}^T, \ldots, u_{k,N}^T]^T = u_{k} = K_{k} x_k,
\end{align}
where $K_k$ is the gain matrix as
\begin{align}
K_{k} = - (B_{k}^T P_{k+1} B_{k} + R_{k})^{-1} B_{k}^T P_{k+1} A_k , \notag
\end{align}
and $P_{k}$ is the recursive matrix as
\begin{align} \label{equ:centralizedc}
P_{k} = & A_k^T (P_{k+1}^{-1} +  B_{k} R_{k}^{-1} B_{k}^T  )^{-1}   A_k
  + Q_{k}  ,
\end{align}
with $P_{M} = Q_{M}$. This centralized controller lacks robustness since it depends heavily on the  central unit. Besides, all input matrices have to be collected and sent to the central processor. When dealing with situations like Example 1 where no central processor is in charge of all agents, this centralized method is infeasible.

\vspace{6pt}
\noindent
\textbf{Framework 2:}  To compensate for the limitation in Framework 1, it is necessary to develop an autonomous strategy for agents to design its own controller, eventually optimizing $J_{M} $ in  (\ref{equ:Optp1}) cooperatively \cite{shoarinejad1999two}. Note that, $J_{M} $ is equivalent to
\begin{align}   \label{equ:JMCOM}
& J_{M}
=   \sum_{i=1}^{N} J_{M,i},
\end{align}
where
\begin{align}  \label{equ:JMi}
J_{M,i}  =   \sum_{k=0}^{M}   x_{k}^T (\pi_{i} Q_{k}) x_{k} + \sum_{k=0}^{M-1} u_{k,i}^T R_{k,i} u_{k,i},
\end{align}
with $\sum_{i=1}^{N} \pi_{i} =1$ and $\pi_{i} > 0$. A typical idea is that each agent optimizes its own cost function $J_{M,i}$. Then, the optimal control law for agent $i$ is derived as follows:
\begin{align} \label{equ:indivc}
u_{k,i}  & = K_{k,i} x_k,
\end{align}
where $K_{k,i}$ is the gain matrix as
\begin{align}
K_{k,i} = - (B_{k,i}^T P_{k+1,i} B_{k,i} + R_{k,i})^{-1} B_{k,i}^T P_{k+1,i} A_k,  \notag
\end{align}
and $P_{k,i}$ is the recursive matrix as
\begin{align}
P_{k,i} = & A_k^T (P_{k+1,i}^{-1} +  B_{k,i} R_{k,i}^{-1} B_{k,i}^T  )^{-1}  A_k + \pi_{i} Q_{k}  , \notag
\end{align}
with $P_{M,i} =  Q_{M} $. Although the control law in (\ref{equ:indivc}) is determined by every agent, due to the possible uncontrollability of $(A_k$, $B_{k,i})$, $P_{k,i}$ might diverge, then leading to the divergence of $u_{k,i} $.
%
%
%
%
%

\noindent
\textbf{Framework 3:} To simultaneously guarantee the stability of the controller and avoid using global information for each agent, an average consensus information fusion strategy has been adopted in \cite{talebi2019distributed}, where the controller for agent $i$ is designed as
\begin{align} \label{equ:consensusc}
u_{k,i}  & =  - R_{k,i}^{-1} B_{k,i}^T {\Theta}_{k+1,i} A_k x_k,
\end{align}
where
\begin{align}
\Psi_{k,i} & = \Gamma_{k,i}^{-1} + N  B_{k,i}  R_{k,i}^{-1} B_{k,i}^T, \notag \\
{\Theta}_{k,i}^{-1} & \longleftarrow \overline{ACF} \longleftarrow \{\Psi_{k,j} , \forall j \in \mathcal{N}_i \} , \notag \\
\Gamma_{k,i} & =  A_k^T {\Theta}_{k,i} A_k + Q_{k}  , \notag
\end{align}
and $\overline{ACF}$ is an iterative consensus function. It should be noted that the above algorithm requires that each agent exchanges information with neighbors infinite times at each control step, requiring a very high communication and computation cost.

The above three frameworks have  limitations in  individuality, stability  and communication-saving. In particular, when the system state $x_{k}$ is not available for all agents in time, these frameworks may fail. To simultaneously overcome the aforementioned drawbacks, this paper   provides a novel suboptimal solution to Problem (\ref{equ:Optp1}) in a fully distributed manner, i.e., each agent designs its controller only by using its own and its neighbors' information.
Specifically, two main issues are focused as follows.

\textbf{1)}: Design a fully distributed controller for each agent based on the LQR problem (\ref{equ:Optp1}).

\textbf{2)}: Analyze the control performance of the proposed controller on both a finite horizon and the infinite horizon.

\section{Main results}\label{sec3}

\subsection{Controller Design} \label{sec3.1}
In this section, a novel class of distributed LQR-based controllers are developed for system (\ref{system_model1}).

First, introduce a series of recursive matrices as
\begin{align}
\label{equ:1overlineP} \bar{P}_{k+1,i} = & (\breve{P}_{k+1,i}^{-1} + B_{k,i} R_{k,i}^{-1} B_{k,i}^{T} )^{-1} , \\
 \label{equ:1P}
P_{k+1,i} = & \Big (\sum_{j \in \mathcal{N}_i }    \omega_{ij}  \bar{P}_{k+1,j}^{-1} \Big )^{-1}  ,    \\
\label{equ:1breveP} \breve{P}_{k,i} = & A_k^T P_{k+1,i} A_k + N Q_{k} ,
\end{align}
where $k = 0, 1,\ldots, M-1$, $\breve{P}_{M,i} = N Q_{M} $, and $\omega_{ij}$ is a coupling gain that satisfies $\omega_{ij} \in (0, \ 1/d_{j}^{out}] $, $j \in \mathcal{N}_i $. It should be noted that $P_{k,i}$ is derived through the backward process. The design of the above recursive matrices is not ad hoc but with intrinsic motivations. On one hand, the iterations of recursive matrices in (\ref{equ:1overlineP}) and (\ref{equ:1breveP}) follow the form of the traditional centralized Ricatti equation in (\ref{equ:centralizedc}). If one sets $\breve{P}_{k,i}=P_{k,i}$ in (\ref{equ:1overlineP}) and (\ref{equ:1breveP}) and omitting (\ref{equ:1P}), the iterative rule of  $\breve{P}_{k,i}$ is the same with that of $P_{k}$ in (\ref{equ:centralizedc}). On the other hand, to address the issue that $(A_k, B_{k,i})$ is uncontrollable, a local fusion strategy (\ref{equ:1P}) is adopted. Compared to the fusion method in \cite{talebi2019distributed}, the one here only demands each agent exchange $ \bar{P}_{k,i}$ with its neighbors once at each step.

After agents obtain these recursive matrices by (\ref{equ:1overlineP})-(\ref{equ:1breveP}), they reverse the communication directions. Then, $\mathcal{G}=(\mathcal{N}, \ \mathcal{E})$ turns to be $\bar{\mathcal{G}}=(\mathcal{N}, \ \bar{\mathcal{E}})$, where $\bar{\mathcal{E}}$ is the same with $\mathcal{E}$ except for the direction. Since agents have the ability to send and receive messages, they only need to exchange the in-neighbor and out-neighbor sets. Hence, it is physically feasible for most of communication links, such as wireless links. Several features of the reversion deserve to be stressed. \textbf{First}, when Assumption \ref{asm:Communication} holds, $\bar{\mathcal{G}}$ is also directed and strongly connected. Thus, this reversion does not change any connectivity of the communication topology.  \textbf{Second}, the reversion only needs to happen once, and the communication graph at every step can be directed.  Hence, it almost increases no communication cost.  \textbf{Moreover}, compared with the undirected graph, the one considered in this paper is more general, since the former is a special case of the latter. Before moving on, Let $\bar{\mathcal{N}}_i $  denote the in-neighborhood set of agent $i$ in $\bar{\mathcal{G}}$, i.e., $\bar{\mathcal{N}}_i = \{j  |  (i, \ j) \in \bar{\mathcal{E}},  \ j \in \mathcal{N}\}$.

Next, a virtual system for agent $i$, $i \in \mathcal{N}$, is introduced as follows:
\begin{align} \label{system_model11}
x_{k+1,i}=  \sum_{j \in \bar{\mathcal{N}}_i } \omega_{ji} \bar{P}_{k+1,i}^{-1}  P_{k+1,j}  A_{k} x_{k,j} + B_{k,i} u_{k,i},
\end{align}
where the initial value $ x_{0,i}$ is chosen as $\frac{1}{N}x_{0}  $.  Then, a useful relation between $x_{k}$ in (\ref{system_model1}) and $x_{k,i} $ in (\ref{system_model11}) can be established as the following lemma.

\begin{lemma} \label{lemma_w}
When $x_{0,i}=\frac{1}{N}x_{0}  $, $x_{k}= \sum_{i=1}^{N} x_{k,i} $ holds for all $ k = 1, \ldots, M-1$.
\end{lemma}

\begin{proof}
It follows from (\ref{system_model1}) that
\begin{align}
\sum_{i=1}^{N} x_{1,i} = & \sum_{i=1}^{N}  \sum_{j \in \bar{\mathcal{N}}_i } \omega_{ji} \bar{P}_{1,i}^{-1}  P_{1,j}  A_{0} x_{0,j} + \sum_{i=1}^{N}  B_{0,i} u_{0,i}  \notag \\
 = & \sum_{i=1}^{N}  \sum_{j \in \mathcal{N}_i } \omega_{ij} \bar{P}_{1,j}^{-1}  P_{1,i}  A_{0} x_{0,i} +   \sum_{i=1}^{N}  B_{0,i} u_{0,i}    \notag \\
 = & \sum_{i=1}^{N} \bigg (  \sum_{j \in \mathcal{N}_i } \omega_{ij} \bar{P}_{1,j}^{-1} \bigg ) P_{1,i}  A_{0} x_{0,i} +   \sum_{i=1}^{N}  B_{0,i} u_{0,i}.  \notag
\end{align}
Then,  substituting (\ref{equ:1P}) into the above equation yields
\begin{align}
\sum_{i=1}^{N} x_{1,i} = & \sum_{i=1}^{N}       A_{0} x_{0,i} +   \sum_{i=1}^{N}  B_{0,i} u_{0,i} = x_{1}.  \notag
\end{align}
By mathematical induction, $x_{k}= \sum_{i=1}^{N} x_{k,i} $ holds for all $k \ge 1$.
\end{proof}

\begin{remark}
System (\ref{system_model1}) is strongly coupled in the sense that its state is affected by all inputs. According to \textbf{Frameworks 1-3}, if this state is directly used to develop a feasible controller, almost all the input matrices are needed for each agent to design the controller gains. To avoid using this global formation, a series of auxiliary vectors $x_{k,i}$ are introduced.
From Lemma \ref{lemma_w}, under only one condition about initial states, the equivalence relation between the state of system (\ref{system_model1}) and the sum of the states of system (\ref{system_model11}) can be established. Based on this, system (\ref{system_model11}) acts as an alternative for system (\ref{system_model1}) for the controller design,  rendering the designing process to be fully distributed.
\end{remark}

Here, each agent needs to collect the information of $x_{0}$ to set $x_{0,i}$. Two basic cases are discussed as follows.

First, consider the case where only some agents have access to $x_{0}$. Without loss of generality, assume that $x_{0}$ is available for agent $i$, $i \in \mathcal{N}_0$, where $ \mathcal{N}_0$ is a subspace of $ \mathcal{N}$. Then, agent $i$, $i \in  \mathcal{N} $, can obtain $x_{0}$ by a fully distributed consensus algorithm as follows:
\begin{align}
\hat{x}_{0,i}^{l} & = x_{0}, & i \in \mathcal{N}_0,   \notag \\
\hat{x}_{0,i}^{l} & = \hat{x}_{0,i}^{l-1} + \frac{1}{N} \sum_{j \in  \bar{\mathcal{N}}_i  }  (  \hat{x}_{0,j}^{l-1}   -   \hat{x}_{0,i}^{l-1}   ), & i \in  \mathcal{N} \setminus \mathcal{N}_0,   \notag
\end{align}
where $l=0,1,\ldots$, and $\hat{x}_{0,i}^{0} = 0$, $i \in  \mathcal{N} \setminus \mathcal{N}_0$.
According to \cite[Lemma 3]{olfati2007consensus}, if Assumption \ref{asm:Communication} holds, $\hat{x}_{0,i}^{l}$ converges to $x_{0}$ exponentially. Hence, each agent can obtain $x_{0}$  quickly.

Second, consider the case where each agent has access to part information of $x_{0}$, denoted by $  H_i x_{0}$ with $H_i \in \mathbb{R}^{r_i \times n}$ for agent $i$. Here, assume that rank($H$) $= n$ with $H = [H_1^T, \ldots, \ H_n^T ]^T$. If this assumption is not satisfied, even the centralized methods with global information will fail to solve problem (2) since  $x_{0}$ cannot be fully achieved \cite[Chapter 4]{bertsekas1995dynamic}.
In this case, agent $i$, $i \in  \mathcal{N}$, obtains $x_{0}$ by a distributed algorithm as follows:
\begin{align}
\bar{x}_{0,ii}^{l} & = H_i x_{0},    \notag \\
\bar{x}_{0,ij}^{l} & = \bar{x}_{0,ij}^{l-1} + \frac{1}{N} \sum_{h \in  \bar{\mathcal{N}}_i    }  (  \bar{x}_{0,hj}^{l-1}   -   \bar{x}_{0,ij}^{l-1}   ),  \  j \neq i, \ j \in \mathcal{N},  \notag
\end{align}
where  $l=0,1,\ldots$, and $\bar{x}_{0,ij}^{0} = 0$, $j \neq i, \ i, j \in \mathcal{N}$. Similar to the above case, agent $i$ can obtain $H_j x_{0}$, $ j \in \mathcal{N}$, quickly. By the same method, the constant matrices $H_j$, $ j \in \mathcal{N}$, can also be received by agent $i$. Hence, agent $i$ can obtain $x_{0}$ by computing $(H^T H)^{-1} H^T \times (H x_{0})$.

For other cases, e.g., only some agents have access to part information of $x_{0}$, agents can combine the distributed algorithms for the two basic cases to collect $x_{0}$.
\vspace{6pt}

Now, a novel controller for agent $i$, $ i \in \mathcal{N}$, is designed as follows:
\begin{align} \label{equ:distc}
u_{k,i} & = K_{k,i}^{dis}  \sum_{j \in \bar{\mathcal{N}}_i } \omega_{ji} \bar{P}_{k+1,i}^{-1}  P_{k+1,j}  A_{k} x_{k,j},
\end{align}
where
\begin{align}  \label{equ:distgain}
 K_{k,i}^{dis} = - ( R_{k,i}  + B_{k,i}^T \breve{P}_{k+1,i} B_{k,i} )^{-1}  B_{k,i}^T \breve{P}_{k+1,i}.
\end{align}
It can be seen that $u_{k,i}$ in (\ref{equ:distc}) is a linear state feedback controller, and the control gain $K_{k,i}^{dis}$ is designed based on the above recursive matrix $\breve{P}_{k+1,i}$.  The whole designing process of the controller for each agent is fully distributed, without using any global information.

Altogether, the controller designing and performing processes are summarized as Algorithm \ref{algorithm1}.

\begin{algorithm}[t]
\caption{ Distributed LQR-based controller designing and performing processes for agent $i$} \label{algorithm1}
\hspace*{0.00in} {\bf 1) Designing process}

\hspace*{0.02in} {\bf Initialization:} $\breve{P}_{M,i} = N Q_{M} $ and $ \omega_{ij}$, $j \in \mathcal{N}_i   $;

\hspace*{0.02in} {\bf for} $ k=M-1;k \ge 0; k - -$, do
\begin{algorithmic}[1]
\State collect $\bar{P}_{k+1,j}$ from agent $j$, $j \in \mathcal{N}_{i} $;

\State calculate  $P_{k+1,i}$ and $\breve{P}_{k,i}$ by (\ref{equ:1overlineP})-(\ref{equ:1breveP});

\end{algorithmic}

\hspace*{0.02in} {\bf end for}

\vspace{10pt}

\hspace*{0.00in} {\bf 2) Performing process}

\hspace*{0.02in} {\bf Initialization:} $x_{0,i} = \frac{1}{N}x_{0}  $;

\hspace*{0.02in} {\bf for} $ k=0;k \leq M; k + +$, do

\begin{algorithmic}[1]
\State collect $x_{k,j}$  from agent $j$, $j \in \bar{\mathcal{N}}_{i}$;

\State calculate $K_{k,i}$ by (\ref{equ:distgain});

\State calculate $u_{k,i}$ by (\ref{equ:distc});

\State calculate $x_{k+1,i}$ by (\ref{system_model11});

\end{algorithmic}

\hspace*{0.02in} {\bf end for}
\end{algorithm}

\subsection{Time-varying systems} \label{sec3.2}
 In this section, the control performance for linear time-varying systems  is analyzed.

\begin{assumption} \label{asm:Invertibility} [Invertibility and boundedness]
 $ A_{k}$ in system (\ref{system_model1}) is non-singular. There exist positive constants $\kappa_{A,1}$, $\kappa_{A,2}$, $\kappa_{B}$, $\kappa_{Q,1}$, $\kappa_{Q,2}$, $\kappa_{R,1}$ and $\kappa_{R,2}$ such that $\kappa_{A,1} \leq \| A_{k} \|_2 \leq \kappa_{A,2}$, $ \| B_{k,i} \|_2 \leq \kappa_{B} $, $\kappa_{Q,1} \leq \| Q_{k} \|_2 \leq \kappa_{Q,2}$ and $\kappa_{R,1} \leq \| R_{k,i} \|_2 \leq \kappa_{R,2}$, $\forall i \in \mathcal{N}$, $\forall k = 0,1,\ldots, M$.
\end{assumption}

\vspace{6pt}

\begin{remark}
The invertibility of $A_{k}$ naturally holds since it is derived by the discretization of continuous-time systems. When $A_{k}$ is singular, one can replace it by any invertible matrix $A_{k,0}$ in its $\epsilon$-neighborhood, i.e., $\|A_{k,0} - A_{k} \|_2 \leq \epsilon $ with $\epsilon$ being a small positive constant. Then, by referring to \cite{zhou1996robust} and \cite{duan2020auto}, the analysis techniques in robust control theory can be applied here directly. Generally, this assumption is mild and reasonable.
\end{remark}

\vspace{6pt}

\begin{proposition}
The computational complexity of Algorithm \ref{algorithm1} for  agent $i$ is $O(d_{i}  n^3 + m_i^3)$, where $d_{i}$, $n$ and $m$ are defined in Section \ref{sec2}.
\end{proposition}

\vspace{6pt}

According to \cite[Chapter 15.3]{cormen2009introduction}, for any matrices $S_1 $, $\ldots$, $S_m \in \mathbb{R}^{n \times n}$, the matrix-chain multiplication problem, i.e.,  $S_1 \times  \cdots \times S_m $, can be solved in $O(n^3)$ time. Hence, it follows from (\ref{equ:1overlineP})-(\ref{equ:distgain}) that Algorithm \ref{algorithm1} can be computed in $O(d_i  n^3 + m_i^3)$ time. Compared with the centralized algorithms in \cite[Chapter 4]{bertsekas1995dynamic}, whose computational complexity is $O(n^3 + (\sum_{i=1}^{N} m_i)^3)$,  the strategy in this paper shows superiority of saving computational resources when $N$ increases.

\vspace{6pt}

Now, for a finite horizon, the stability and the control performance of the proposed controller are summarized as the following theorems.

\vspace{6pt}

\begin{theorem}\label{thm1}
If Assumptions \ref{asm:Communication}, \ref{asm:Joint} and \ref{asm:Invertibility} hold,  for any finite positive integer $M$, the controller parameters $P_{k,i}$, $ \breve{P}_{k,i}$ and $\bar{P}_{k,i}$ are uniformly bounded.
\end{theorem}

\vspace{6pt}

The proof of Theorem \ref{thm1} is given in Appendix \ref{proof1}.

Actually, $P_{k,i}$, $ \breve{P}_{k,i}$ and $\bar{P}_{k,i}$ in (\ref{equ:1overlineP})-(\ref{equ:1breveP}) are not Ricatti recursive matrices that facilitate the optimal LQR design. Instead, they are introduced to act as control indicators, i.e., to evaluate the control performance.  Moreover, Theorem \ref{thm1} reveals that the gain $K_{k,i}^{dis}$ in (\ref{equ:distc}) is uniformly bounded, which guarantees the feasibility of the proposed controller.

\vspace{6pt}

\begin{theorem}\label{thm2}
Considering system (\ref{system_model1}) with inputs designed by (\ref{equ:distc}), if Assumptions \ref{asm:Communication},  \ref{asm:Joint} and \ref{asm:Invertibility} hold,  for any finite positive integer $M$, the optimized cost function $J_M$ is upper bounded by
 \begin{align} \label{equ:optzero1}
& J^{bound}_{M} = \frac{1}{N^2} \sum_{i=1}^{N}  x_0^T  \breve{P}_{0,i} x_0.
\end{align}
\end{theorem}
\vspace{6pt}

The proof of Theorem \ref{thm2} is given in Appendix \ref{proof2}. In this proof, one condition about $w_{ij}$ needs to be satisfied, i.e., $\omega_{ij} \in (0, \ 1/d_{j}]  $, which plays an essential role in guaranteeing the boundedness of the performance index.

\vspace{6pt}
\begin{remark}
Since $\breve{P}_{0,i}$ in (\ref{equ:optzero1}) is uniformly upper bounded, $J^{bound}_{M}$ has a uniform upper bound. Moreover, Theorem \ref{thm2} reveals two results. First, the control performance index defined in Problem (\ref{equ:Optp1}) is guaranteed by the proposed algorithm. Second, the state of system (\ref{system_model1}) is upper bounded.
\end{remark}

\vspace{6pt}
\begin{remark}
Compared with the consensus-based  controller in \cite{talebi2019distributed},  Algorithm \ref{algorithm1} achieves a better trade-off between the control performance and the communication cost. First, the performance index is upper bounded by a closed-form expression only concerning the system initial value and the system matrices. Second, at each designing and performing step, each agent only needs to deliver its information to neighbors once, rather than infinite times in \cite{talebi2019distributed}.  Therefore, in this paper, a suboptimal solution is obtained with much less communication consumption.
\end{remark}

 \vspace{6pt}
From Theorem \ref{thm2}, the performance index $J_{M}$ is upper bounded by $ \sum_{i=1}^{N}   x_{0}^T (\breve{P}_{0,i} /N^2)x_{0}$, where $\breve{P}_{0,i}$ is affected by $\omega_{ij}$, $j \in \mathcal{N}_i $, from (\ref{equ:1P}). To further improve the control performance for arbitrary  initial states, it is expected to minimize $ \lambda_{max} (  \breve{P}_{0,i} ) \ (= \| \breve{P}_{0,i} \|_2)$ with respect to $\omega_{ij}$ for agent $i$. Hence, it follows from (\ref{equ:1breveP}) that $\omega_{ij}$ at step $k = 1$ can be obtained by solving the following optimization problem:
\begin{align}
   \min_{ \omega_{ij}, \ j \in \mathcal{N}_i  } & \  \|  A_0^T P_{1,i} A_0 \|_2,   \notag  \\
\text{ s.t.}  \  (\ref{equ:1P})  \ \&   & \ \omega_{ij}  \in (0, \ 1/d_{j}^{out}].  \notag
\end{align}
The cost function in the above problem is also affected by $ \bar{P}_{1,j}$ from (\ref{equ:1P}), further by $P_{2,j}$ from (\ref{equ:1overlineP}) and (\ref{equ:1breveP}). Note that when $P_{2,i}$ is a scalar, it can be verified that a smaller $ P_{2,i} $ leads to a smaller cost function. In this case, one can minimize $ P_{2,i} $ with respect to $\omega_{ij}$ at step $k=2$.
By borrowing this idea to the matrix case, $\omega_{ij}$ at step $k \in \{2,\ldots, M\}$, denoted by $\omega_{ij,k}$,  is computed for agent $i$ by solving the optimization problem as follows:
\begin{align}
   \min_{ \omega_{ij,k}, \ j \in \mathcal{N}_i  } & \  \| P_{k,i} \|_2,    \notag \\
\text{ s.t.}  \  (\ref{equ:1P})  \ \&   \ \omega_{ij,k}  & \in (0, \ 1/d_{j}^{out}].  \notag
\end{align}
Many existing methods, such as the quasi-Newton method \cite{BartholomewBiggs2008Nonlinear}, can be adopted to solve the above problem.

The optimal $\| \breve{P}_{0,i} \|_2$ is concerned with $\omega_{ij,k}$, $\forall i \in \mathcal{N} $,  $\forall j \in \mathcal{N}_i $, $ \forall k = 1, \ldots, M$, where weights $\omega_{ij,k}$ are strongly coupled with each other. It is too computationally time-consuming to obtain the optimal solution. In this paper, this complex problem is approximately decoupled such that each agent only needs to solve a local optimization problem with much lower computational complexity. Besides,  the solving process of $\omega_{ij}$ is backward from $k=M$ to $k=1$, which is fully consistent with the derivation process of $P_{k,i}$ in (\ref{equ:1P}). Moreover, a closed-form relation between $\omega_{ij}$ and $P_{k,i}$ will be established in the next subsection.

\subsection{Time-invariant systems}\label{sec3.3}

In this section, the control performance for linear time-invariant systems  is analyzed.

\begin{assumption} \label{asm:Time-invariant} [Time-invariant]
The system (\ref{system_model1}) reduces to a   time-invariant system, i.e., $A_{k} = A$, $B_{k,i} = B_{i}$, $Q_{k} = Q$ and $R_{k,i} = R_{i}$, $\forall i \in \mathcal{N} $.
\end{assumption}

The results in Theorems \ref{thm1} and \ref{thm2} are extended to the case with the infinite horizon as follows:

\begin{theorem} \label{thm3}
If Assumptions \ref{asm:Communication},  \ref{asm:Joint},  \ref{asm:Invertibility} and \ref{asm:Time-invariant} hold, and $M= \infty$ in (\ref{equ:Optp1}),  the controller parameters $P_{k,i}$, $ \breve{P}_{k,i}$ and $\bar{P}_{k,i}$ converge to
\begin{align}
 P_{i} =   \Big ( \sum_{j \in \mathcal{N}_i } &  \omega_{ij}  \big ( (A^T P_{j} A + N Q)^{-1}  + B_{j} R_{j} B_{j}^{T} \big )   \Big )^{-1} ,  \notag  \\
\qquad \qquad & \breve{P}_{i} =  A^T P_{i} A + N Q, \notag \\
\qquad \qquad & \bar{P}_{i} =  (\breve{P}_{i}^{-1} + B_{i} R_{i}^{-1} B_{i}^{T} )^{-1} . \notag
\end{align}
\end{theorem}

The proof of Theorem \ref{thm3} is given in Appendix \ref{proof3}.

For time-invariant cases, matrices $P_{i}$, $\breve{P}_{i}$ and $\bar{P}_{i}$, $\forall i \in \mathcal{N}$, can be given in the forms of a series of Ricatti-like equations. By denoting $P_{aug}^{-1}= [P_{1}^{-1}$, $\ldots$, $P_{N}^{-1}]^T$, one has
\begin{align}
 P_{aug}^{-1} = (\Pi \otimes I_n) &      \left[{
\begin{array}{*{20}{c}}
  {(A^T P_{1} A + N Q)^{-1}  + B_{1} R_{1} B_{1}^{T}} \\
  {\vdots}  \\
    {(A^T P_{N} A + N Q)^{-1}  + B_{N} R_{N} B_{N}^{T}} \\
  \end{array} }\right],  \notag
\end{align}
where $\Pi = [ \omega_{ij}]_{N \times N}$. It can be found that $P_i$, $ i \in \mathcal{N}$, are strongly coupled with each other from the above equation.  Even so, $P_{i}$, $\breve{P}_{i}$ and $\bar{P}_{i}$ can be computed by  (\ref{equ:1overlineP})-(\ref{equ:1breveP}) after sufficient fully distributed iterations, similarly to the Riccati matrix in the standard LQR control.  Moreover, in the standard infinite horizon LQR control, $(\sqrt{Q}, \ A)$ needs to be detectable to stabilize the Ricatti equation \cite[Corollary 13.8]{zhou1996robust}, where $Q = (\sqrt{Q})^T \sqrt{Q}$. Since $(\sqrt{Q}, \ A)$ must be observable when $Q > 0$, this condition is always satisfied in this paper.

\begin{theorem}\label{thm4}
Considering system (\ref{system_model1}) with inputs designed by (\ref{equ:distc}), if Assumptions \ref{asm:Communication}, \ref{asm:Joint},  \ref{asm:Invertibility} and \ref{asm:Time-invariant} hold, and $M= \infty$ in (\ref{equ:Optp1}), the optimized cost function $J_M$ is upper bounded by
 \begin{align} \label{equ:optzero2}
& J^{bound}_{\infty} \triangleq \frac{1}{N^2} \sum_{i=1}^{N} x_0^T  \breve{P}_{i} x_0.
\end{align}
\end{theorem}

The proof of Theorem \ref{thm4} is given in Appendix \ref{proof4}.

The convergence of the system state can be further ensured as the following corollary.

\begin{corollary} \label{corollary1}
Considering system (\ref{system_model1}) with inputs designed by (\ref{equ:distc}), if Assumptions \ref{asm:Communication}, \ref{asm:Joint},  \ref{asm:Invertibility} and \ref{asm:Time-invariant} hold, and $M= \infty$ in (\ref{equ:Optp1}), the state of system (\ref{system_model1}) converges to zero.
\end{corollary}

The proof of Corollary \ref{corollary1} is given in Appendix \ref{corollary1proof}.

In the above results, the feedback states are chosen as the virtual ones $x_{k,i}$ defined in (\ref{system_model11}), which is based on the initial states and system parameters. To formulate the closed-loop form, it is preferable to utilize the real-time states properly. For this purpose, take the finite horizon case for example. First, the time line is divided into the following intervals, i.e.,
 \begin{align}
 \{0,1,\ldots,M\} \longrightarrow    \left \{ {
\begin{array}{*{20}{c}}
  { \{0,1,\ldots,\overline{N} \},} \\
  {\{\overline{N}+1,\overline{N},\ldots,2\overline{N}\},} \\
  {\vdots}  \\
   {\{\xi \overline{N}+1,\xi \overline{N}+1,\ldots,M\},}  \\
  \end{array} }\right \}  \notag   ,
\end{align}
where $M= \xi \overline{N} + \overline{N}_{re}$ with $\xi$,  $\overline{N}$ and $\overline{N}_{re}$ being non-negative integers satisfying $\overline{N} \ge N+L$ and $0 \leq \overline{N}_{re} < \overline{N}$. Then, during the time interval $\{h\overline{N}+1,  h \overline{N}+2, \ldots, (h+1)\overline{N}\}$, $h=0$, $\ldots$, $\xi$, define the virtual system as ({\ref{system_model11}) with the initial value being $\frac{1}{N}x_{h\overline{N}}$. Besides, set $\breve{P}_{(h+1)\overline{N},i}$ = $ N \overline{Q}_{(h+1)\overline{N}} $ with $\overline{Q}_{(h+1)\overline{N}} = Q_{(h+1)\overline{N}} + \frac{1}{N}\sum_{i=1}^{N}  A^T_{(h+1)\overline{N}} P_{(h+1)\overline{N}+1,i} A_{(h+1)\overline{N}} $. Other parts in Algorithm \ref{algorithm1} remain valid. Similarly to the result in Theorem \ref{thm2}, the control performance can be ensured.

\section{Simulation}\label{sec4}

In this section, two numerical simulation results are provided to illustrate the effectiveness of Algorithm \ref{algorithm1}.

{\bf{Part 1}}: A modified control task with 8 UAVs is considered, which is borrowed from   \cite{xu2019decentralized}.  The system model of this setting is described as (\ref{system_model1}), and the state vector $x_k=[x_{k}(1), \ \ldots, x_{k}(8)]^T$ denotes the positions (errors) of all nodes. The communication topology of nodes is shown in Fig. \ref{f:communication_topology}. It is expected that the positions   converge to zero, with inputs developed and performed in a fully distributed way.
\begin{figure}[!htb]
\center
\subfigure{{\includegraphics[scale=0.45]{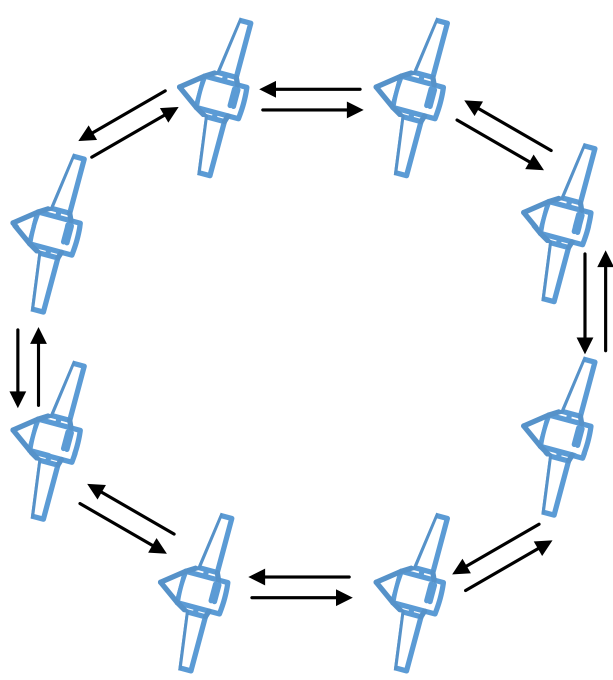}}}
\caption{The communication topology graph.} \label{f:communication_topology}
\end{figure}
In this example, the state matrix is set as $A_{k} = I_8 + [0_{7 \times 1}, \ \delta \times I_8; \  0, \ 0_{1 \times 7} ]$,  where the non-diagonal  element $\delta(=0.02)$  denotes the intrinsic coupling between nodes.  The input matrices are set as $  [{B_{k,1}}, \ {B_{k,2}}, \ {B_{k,3}}, \ {B_{k,4}}, \ {B_{k,5}}, \ {B_{k,6}}, \ {B_{k,7}}, \ {B_{k,8}}]  =0.5 \times I_8  $. Besides, the initial state is chosen as $x_{0}= [x_{0}(1), \ \ldots, x_{0}(8)]^T $, where $x_{0}(i)=3+2 \times i$, $\forall i=1$, $\ldots$, $8$. The weights in (\ref{equ:Optp1}) are selected as $Q_{k}= 20 \times I_8$  and $R_{k,i}=30$, $\forall i=1$, $\ldots$, $8$, $\forall k=0$, $\ldots$, $M$, where $M=120$.


\begin{figure}[!htb]
\center
\subfigure{{\includegraphics[scale=0.45]{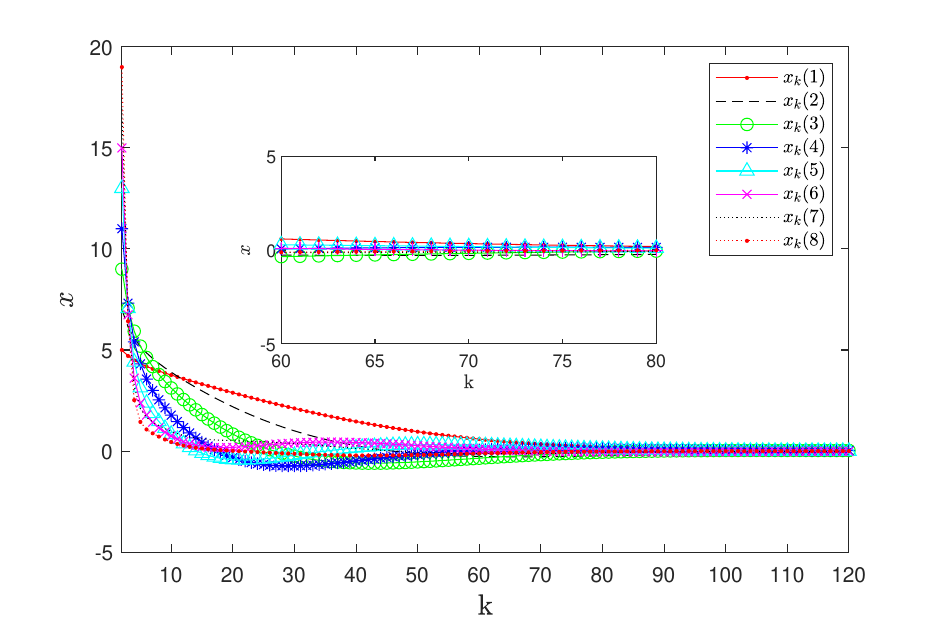}}}
\caption{The states of all nodes in  Part 1.} \label{f:state}
\end{figure}


\begin{figure}[!htb]
\center
\subfigure{{\includegraphics[scale=0.45]{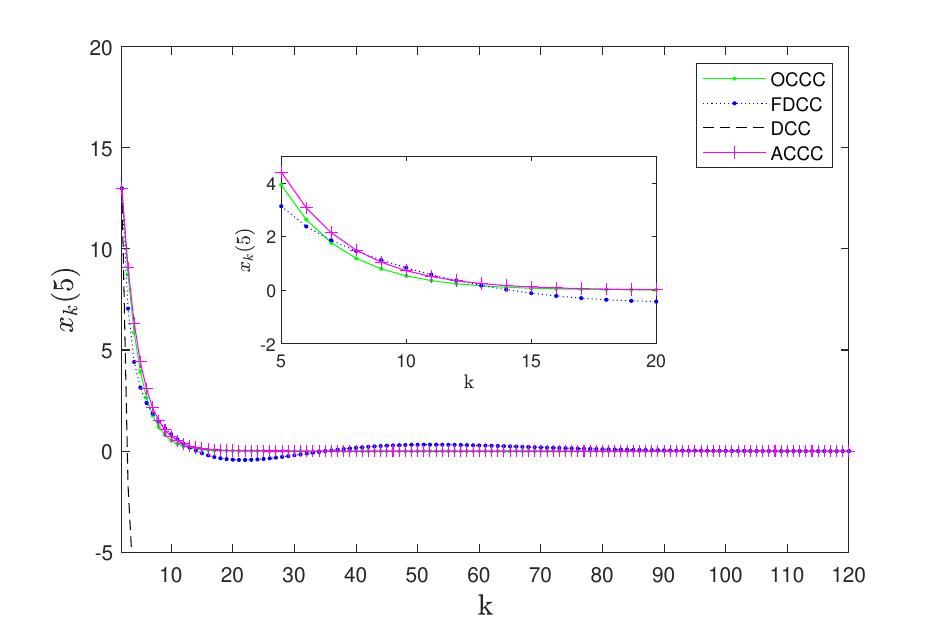}}}
\caption{The values of $x_{k}(5)$ by different controllers.} \label{f:com}
\end{figure}

First, by applying Algorithm \ref{algorithm1} with the above parameters,  numerical simulation results are shown in Fig. \ref{f:state}, where the states of 8 nodes are plotted. It can be seen that the states converge, which verifies the validity of results in Theorems \ref{thm1} and \ref{thm2}.

Then, to show the superiority of the proposed fully distributed cooperative controller (FDCC), comparative results with some typical LQR control strategies are provided, including the optimal centralized cooperative controller (OCCC) with full system information in  \cite[Chapter 4]{bertsekas1995dynamic}, the decentralized cooperative controller (DCC) without information exchange in \cite{shoarinejad1999two}, and the average consensus-based cooperative controller (ACCC) with 12 times communication at every control step in \cite{talebi2019distributed}. Without loss of generality, the states of node 5 by the above controllers are shown in Fig.~\ref{f:com}, where the state trajectory by FDCC is close to that by OCCC.
Moreover, the optimal cost by OCCC is $8.3 \times 10^4$ while the suboptimal cost by FDCC is $11.1 \times 10^4$, which indicates that the proposed controller in this paper guarantees a satisfactory control performance even though it is equipped with much less system information and communication resources.

{\bf{Part 2}:} To illustrate the effectiveness of the proposed controller for time-varying multi-input systems with strong coupling, an example with 50 nodes, modified from \cite{1605401}, is provided. The state matrix is set as
\begin{align}
   A_k    =   1.02 \times I_{50} - 0.01  \times   \mathbb{1}_{50} \times \mathbb{1}_{50}^T 
,  \notag
\end{align}
whose non-diagonal elements denote the coupling between nodes. It is worth noting that $A_k$ is not Schur stable. Besides, the time-varying input matrices $B_{k,i}$, $i = 1$, $\ldots$, $50$, are chosen as
\begin{align}
  B_{k,i}
  =   \begin{cases}
     [\mathbb{0}_{i-1}, \ 2 + \text{cos}(i \times k), &   \mathbb{0}_{50-i}  ]^T,   \   \text{if $ i = 5 h + 1, $ } \\
    [\mathbb{0}_{i-1}, \ 2 + \text{sin}(i \times k), &  \mathbb{0}_{50-i}  ]^T,  \  \text{if $ i = 5 h +2, $ } \\
    [\mathbb{0}_{i-1}, \ 2 ,   & \mathbb{0}_{50-i}  ]^T,   \   \text{if $ i = 5 h + 3, $ } \\
   [\mathbb{0}_{i-1}, \  2 - \text{cos}(i \times k), &   \mathbb{0}_{50-i}  ]^T,  \   \text{if $ i = 5 h + 4, $ } \\
    [\mathbb{0}_{i-1}, \ 2 - \text{sin}(i \times k),  &   \mathbb{0}_{50-i}  ]^T,   \  \text{if $ i = 5 h + 5, $ } \\
  \end{cases} \notag
\end{align}
for all $h = 0 $, $\ldots$, $9$. The initial state is set as $x_{0}= [x_{0}(1), \ \ldots, x_{0}(50)]^T $, where $x_{0}(i)=30+0.1 \times i$, $ i=1$, $\ldots$, $50$.
The communication topology graph of nodes is a cycle, which is similar to Fig.~\ref{f:communication_topology}.  Other parameters are the same with {\bf{Part 1}}. Without loss of generality, the states of  agents {\it 21-25} are illustrated. As shown in Fig.~\ref{f:statepart2}, all states converge to a small value, which shows the effectiveness of Algorithm \ref{algorithm1} for time-varying and strongly coupled systems.
\begin{figure}[!htb]
\center
\subfigure{{\includegraphics[scale=0.45]{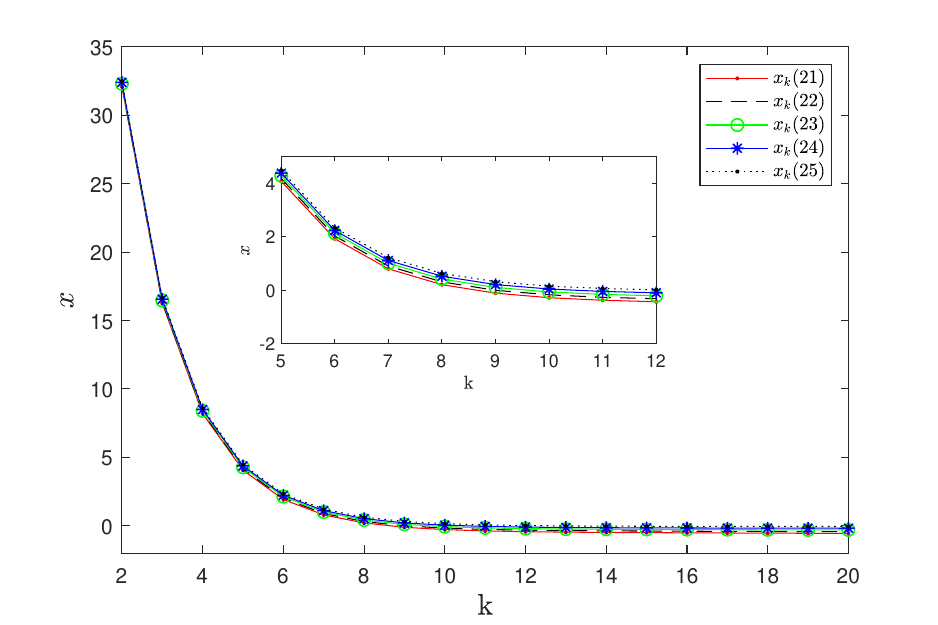}}}
\caption{The states of  agents {\it 21-25} in  Part 2.} \label{f:statepart2}
\end{figure}

\section{Conclusion}\label{sec5}
This paper investigated a cooperative control problem for multi-input systems, where each input was determined by an agent in a network. Particularly, the heterogeneous and time-varying input matrices were considered to be local information for agents. Based on  a novel fully distributed information fusion strategy,  a cooperative controller has been proposed, with an LQR control performance being ensured. For each agent, only one-time information exchange with its neighbors was required at every step. Hence, a better trade-off between the control performance and the communication cost could be achieved. Regarding future works, we will apply the theoretical results to physical systems.

\appendices

\section{Proof of Theorem \ref{thm1}}\label{proof1}

Without loss the generality,  it suffices to prove that $P_{k,i}$ is uniformly bounded, i.e., there exist two positive constants $\rho_1$ and $\rho_1$ such that $\rho_1 I_n \leq P_{k,i} \leq \rho_2 I_n $ ($\rho_2^{-1} I_n \leq P_{k,i}^{-1} \leq \rho_1^{-1} I_n$), $\forall i \in \mathcal{N} $, $\forall k = 0,1,\ldots,M$. In the following, this proof is divided into two parts: 1) $M > \overline{N}$;  2) $M \leq  \overline{N}$, where $\overline{N} = L +N$ and $L$ is given in Assumption \ref{asm:Joint}.

\textbf{1) $M \leq  \overline{N}$:} Since $\overline{N}$ is a given positive constant, there must exist two positive constants $\rho_{1,1}$ and $\rho_{2,1}$ such that $\rho_{1,1} I_n \leq P_{k,i} \leq \rho_{2,1} I_n $, $\forall i \in \mathcal{N} $, $\forall k = 0,1,\ldots,M$.

\textbf{2) $M >  \overline{N}$:} First, the lower uniform boundedness of $P_{k,i}$ is proved, i.e., there exists a positive constant $\rho_{1,2}$ such that $ P_{k,i}^{-1} \leq \rho_{1,2}^{-1} I_n$. If Assumption \ref{asm:Invertibility} holds,   one has
\begin{align}
P_{k,i}^{-1} = &  \sum_{j \in \mathcal{N}_i }  \omega_{ij}  \bar{P}_{k,j}^{-1} \notag \\
 = &  \sum_{j \in \mathcal{N}_i }\omega_{ij} (\breve{P}_{k+1,j}^{-1} + B_{k,j} R_{k,j}^{-1} B_{k,j}^{T} )   \notag \\
 \overset{a}{<} &  \sum_{j \in \mathcal{N}_i } \omega_{ij} (Q_{k}^{-1}/N + B_{k,j} R_{k,j}^{-1} B_{k,j}^{T} )   \notag \\
 \leq &  \sum_{j \in \mathcal{N}_i } \omega_{ij} (1/(N\kappa_{Q,1}) + \kappa_{B}^2/\kappa_{R,1} ) I_n    \notag \\
 \leq &   n_{m}/\kappa_{Q,1} I_n + n_{m} N \kappa_{B}^2/\kappa_{R,1} I_n,  \notag
\end{align}
where $`` \overset{a}{<} " $ is derived from (\ref{equ:1breveP}), and $ n_{m}=\max_{i,j \in \mathcal{N}} \omega_{ij} $. By denoting  $\rho_{1,2}^{-1} \triangleq ( n_{m}/\kappa_{Q,1} + n_{m} N \kappa_{B}^2/\kappa_{R,1} )^{-1}$, one has $ \rho_{1,2} I_n \leq P_{k,i}$,  $\forall i \in \mathcal{N} $, $\forall k = 0,1,\ldots,M$.

Then, the upper uniform boundedness of $P_{k,i}$ is proved, i.e., there exists a positive constant $\rho_{2,2}$ such that $ P_{k,i}^{-1} \ge \rho_{2,2}^{-1} I_n$. It follows from (\ref{equ:1breveP}) and Assumption \ref{asm:Invertibility}   that
\begin{align}
& A_k^T P_{k+1,i} A_k + N Q_{k} \notag \\
\leq & \{1+ \lambda_{max}( N Q_{k})/\lambda_{min}( A_k^T P_{k+1,i} A_k )\}  A_k^T P_{k+1,i} A_k  \notag \\
 \leq & \{1+  N \kappa_{Q,2} \kappa_{A,1}^{-2} \rho_1^{-1}\}  A_k^T P_{k+1,i} A_k,  \notag
\end{align}
where $\rho_1=\min \{\rho_{1,1}, \ \rho_{1,2}\}$. Hence, there must exist a positive constant $\beta \in (0, \ 1)$ such that $(A_k^T P_{k+1,i} A_k + N Q_{k})^{-1} \ge \beta (A_k^T P_{k+1,i} A_k)^{-1} $. Further,  from (\ref{equ:1overlineP})-(\ref{equ:1breveP}), one has
 \begin{align}
& P_{k,i}^{-1} \notag \\
 = &  \sum_{j \in \mathcal{N}_i } \omega_{ij} \Big ((A_k^T P_{k+1,j} A_k + N Q_{k})^{-1} + B_{k,j} R_{k,j}^{-1} B_{k,j}^{T} \Big ) \notag \\
\ge &  \sum_{j \in \mathcal{N}_i } \omega_{ij} \Big (\beta A_k^{-1} P_{k+1,j}^{-1} A_k^{-T} + B_{k,j} R_{k,j}^{-1} B_{k,j}^{T} \Big ),  \notag
\end{align}
which reveals a relation between $P_{k,i}^{-1}$ and $P_{k+1,j}^{-1}$, $j \in \mathcal{N}_i $. By mathematical induction, one has
\begin{small}
\begin{align}
 & P_{k,i}^{-1}   \ge   \beta \sum_{j = 1}^N n_{ij,2}  \Phi_{k,k}^{-1} B_{k+1,j} R_{k+1,j}^{-1} B_{k+1,j}^{T} \Phi_{k,k}^{-T}   \notag \\
 +&  \sum_{j = 1}^N   a_{ij} \omega_{ij}  B_{k,j} R_{k,j}^{-1} B_{k,j}^{T}  + \beta^2 \sum_{j = 1}^N n_{ij,2} \Phi_{k+1,k}^{-1} P_{k+2,j}^{-1} \Phi_{k+1,k}^{-T}  \notag \\
 \ge & \ \cdots \notag \\
 \ge &  \sum_{h = 0}^{\overline{N}-1} \beta^{h} \sum_{j = 1}^N n_{ij,h+1} \Phi_{k+h-1,k}^{-1} B_{k+h,j} R_{k+h,j}^{-1} B_{k+h,j}^{T}  \Phi_{k+h-1,k}^{-T}  \notag \\
  + & \sum_{j = 1}^N \beta^{\overline{N}}  n_{ij,\overline{N}} \Phi_{k+\overline{N}-1,k}^{-1} P_{k+\overline{N},j}^{-1} \Phi_{k+\overline{N}-1,k}^{-T}  \notag \\
> & \sum_{h = N}^{\overline{N}-1} \beta^{h} \sum_{j = 1}^N n_{ij,h+1}  \Phi_{k+h-1,k}^{-1} B_{k+h,j} R_{k+h,j}^{-1} B_{k+h,j}^{T}  \Phi_{k+h-1,k}^{-T}  , \notag
\end{align}
\end{small}
where $n_{ij,h}=[\Pi^h]_{ij}$, where $\Pi = [  \omega_{ij}]_{N \times N}$ and $\Pi^h$ means that $\Pi$ multiplies itself $h$ times. It follows from Lemma \ref{lemma0} that  $n_{ij,h}>0$ for all $h \ge N$. Then, by denoting $n_{n}=\min_{\forall i,j \in \mathcal{N}, h=N,\ldots,\overline{N}-1}\{n_{ij,h}\}$, one further has
\begin{align}
 P_{k,i}^{-1}
> & n_{n} \beta^{\overline{N}-1} \sum_{h = N}^{\overline{N}-1}  \Phi_{k+h-1,k}^{-1} B_{k+h} R_{k+h}^{-1} B_{k+h}^{T}  \Phi_{k+h-1,k}^{-T} .  \notag
\end{align}

According to the joint controllability in Assumption \ref{asm:Joint}, one has
\begin{align}
& \sum_{h = N}^{\overline{N}-1}  \Phi_{k+h-1,k}^{-1} B_{k+h} R_{k+h}^{-1} B_{k+h}^{T}  \Phi_{k+h-1,k}^{-T}  \notag \\
= & \sum_{h = N}^{\overline{N}-1}  \Phi_{k+h-1,k}^{-2} \Phi_{k+h-1,k} B_{k+h} R_{k+h}^{-1} B_{k+h}^{T}  \Phi_{k+h-1,k}^{T} {\Phi_{k+h-1,k}^{-2T}} \notag \\
\ge & \kappa_{A,2}^{-4 \overline{N}}  \sum_{h = N}^{\overline{N}-1} \Phi_{k+h-1,k} B_{k+h} R_{k+h}^{-1} B_{k+h}^{T}  \Phi_{k+h-1,k}^{T} \notag \\
\ge & \kappa_{A,2}^{-4 \overline{N}} \eta I_n. \notag
\end{align}
Thus,  one has $P_{k,i}^{-1}  \ge n_{n} \beta^{\overline{N}-1} \kappa_{A,2}^{-4 \overline{N}} \eta I_n \triangleq  \rho_{2,2}^{-1}  I_n $.
Now, by denoting $\rho_1=\min \{\rho_{1,1}, \ \rho_{1,2}\}$ and $\rho_2=\max \{\rho_{2,1}, \ \rho_{2,2}\}$, one further has  $\rho_1 I_n \leq P_{k,i} \leq \rho_2 I_n$, which indicates that $P_{k,i}$ is uniformly bounded.

\section{Proof of Theorem \ref{thm2}}\label{proof2}
 First, define a function $\hat{J}_{M,i}$ for agent $i$, as
\begin{align} \label{equ:Hat_J_i}
  \hat{J}_{M,i}   \triangleq \sum_{k=0}^{M}   x_{k,i}^T  (N Q_{k}) x_{k,i} + \sum_{k=0}^{M-1} u_{k,i}^T R_{k,i} u_{k,i}.
\end{align}
Then, one has
\begin{align}  \label{equ:Hat_J}
  &   \sum_{i=1}^{N} \hat{J}_{M,i}  \triangleq  \hat{J}_{M}   \notag \\
    = & \sum_{i=1}^{N} \bigg ( \sum_{k=0}^{M}   x_{k,i}^T  (N Q_{k}) x_{k,i} + \sum_{k=0}^{M-1} u_{k,i}^T R_{k,i} u_{k,i} \bigg ) \notag \\
    = &  \sum_{k=0}^{M} \sum_{i=1}^{N}  x_{k,i}^T  (N Q_{k}) x_{k,i} + \sum_{k=0}^{M-1} u_{k}^T R_{k} u_{k}.
\end{align}
Notice that $x_{k}= \sum_{i=1}^{N} x_{k,i} $ holds for all $k \ge 1$ from Lemma \ref{lemma_w}. Thus, it follows from Lemmas  \ref{lemma1} and \ref{lemma_w} that
\begin{align}
   x_{k}^T Q_k  x_{k} =  &  \Big ( \sum_{i=1}^{N} x_{k,i} \Big )^T Q_k \Big ( \sum_{i=1}^{N} x_{k,i} \Big )
   \leq      \sum_{i=1}^{N} N x_{k,i}^T  Q_{k} x_{k,i}.  \notag
\end{align}
where $x_{k,i}$ and $Q_k$ correspond to $\alpha_i$ and $P$ in Lemma \ref{lemma1}, respectively.

Hence, $\hat{J}_{M}$ in (\ref{equ:Hat_J}) satisfies
\begin{align} \label{equ:Hat_Jleq}
  & \hat{J}_{M} \ge    \sum_{k=0}^{M} x_{k}^T Q_k  x_{k} + \sum_{k=0}^{M-1} u_{k}^T R_{k} u_{k} = J_{M},
\end{align}
which indicates that $\hat{J}_{M}$ is an upper bound of $J_{M}$. Further, the problem (\ref{equ:Optp1}) can be relaxed as
\begin{align} \label{equ:Optp2}
   &   \min_{\stackrel{u_{k,i}, i \in \mathcal{N}} {k=0, \ldots, M-1 }}  \hat{J}_{M},    \\
 s.t. \quad & x_{k+1}  = A_{k} x_k + \sum_{i=1}^{N} B_{k,i} u_{k,i}  . \notag
\end{align}
Before proceeding, denote an auxiliary function $\hat{J}_{h}$ as
\begin{align} \label{equ:Hat_J_h}
  \hat{J}_{h}  \triangleq   \sum_{i=1}^{N} \hat{J}_{h,i},
\end{align}
where
\begin{align} \label{equ:Hat_J_hi}
  \hat{J}_{h,i}  \triangleq \sum_{k=0}^{h}   x_{k,i}^T  (N Q_{k}) x_{k,i} + \sum_{k=0}^{h-1} u_{k,i}^T R_{k,i} u_{k,i}.
\end{align}

Next, it follows from (\ref{equ:Hat_J_i})  and the definition of $P_{M,i}$ below (\ref{equ:1overlineP}) that
\begin{align} \label{equ:Hat_J_mhi}
  \hat{J}_{M,i}  =  \hat{J}_{M-1,i} + x_{M,i}^T  \breve{P}_{M,i} x_{M,i} +  u_{M-1,i}^T R_{M-1,i} u_{M-1,i}.
\end{align}
Now, consider the last two terms of the above equation, i.e., $ x_{M,i}^T  \breve{P}_{M,i} x_{M,i}$ and $u_{M-1,i}^T R_{M-1,i} u_{M-1,i}$.  By directly substituting (\ref{system_model11}) and (\ref{equ:distc}) into these two terms, one has
\begin{align}
 & x_{M,i}^T  \breve{P}_{M,i} x_{M,i} + u_{M-1,i}^T R_{M-1,i} u_{M-1,i} \notag \\
  =  & \Big ( \sum_{j \in \bar{\mathcal{N}}_i } \omega_{ji} \bar{P}_{M,i}^{-1}  P_{M,j}  A_{M-1} x_{M-1,j}  \Big )^T \Big \{ (K_{M-1,i}^{dis})^T   \times   \notag \\
   & R_{M-1,i} (* )  + (  I_n + B_{M-1,i}  K_{M-1,i}^{dis} )^T  \breve{P}_{M,i} (* )  \Big \} \Big ( * \Big ). \notag
\end{align}
It follows from  (\ref{equ:distgain})  that
\begin{align}
  &   (K_{M-1,i}^{dis})^T R_{M-1,i}  (* ) + (  I_n + B_{M-1,i}  K_{M-1,i}^{dis} )^T  \breve{P}_{M,i} (  * )    \notag \\
 = & \  \breve{P}_{M,i}  +    \breve{P}_{M,i} B_{M-1,i}  K_{M-1,i}^{dis}  + (K_{M-1,i}^{dis})^T B_{M-1,i}^T \breve{P}_{M,i}  \notag  \\
 & + ( K_{M-1,i}^{dis})^T ( B_{M-1,i}^T  \breve{P}_{M,i} B_{M-1,i} + R_{M-1,i}) K_{M-1,i}^{dis}
 \notag \\
  \overset{\text{b}}{=} & \ \breve{P}_{M,i}   - \breve{P}_{M,i} B_{M-1,i} ( R_{M-1,i}  + B_{M-1,i}^T \breve{P}_{M,i} B_{M-1,i} )^{-1}  \notag \\
 & \times   B_{M-1,i}^T  \breve{P}_{M,i}   \notag \\
  \overset{\text{c}}{=} & \ ( \breve{P}_{M,i}^{-1} + B_{M-1,i} R_{M-1,i}^{-1} B_{M-1,i}^T )^{-1}
 =  \bar{P}_{M,i},   \notag
\end{align}
where $ `` \overset{\text{b}}{=}"$  is derived by substituting the expression of $K_{M-1,i}^{dis}$ in (\ref{equ:distgain}), and $ `` \overset{\text{c}}{=}"$  is derived based on Woodbury matrix identity \cite{woodbury1950inverting}, \cite[Lemma 1]{duan2020auto}.

Hence, $x_{M,i}^T  \breve{P}_{M,i} x_{M,i} + u_{M-1,i}^T R_{M-1,i} u_{M-1,i}$ in (\ref{equ:Hat_J_mhi}) satisfies
\begin{align} \label{motivation_wij}
 & x_{M,i}^T  \breve{P}_{M,i} x_{M,i} + u_{M-1,i}^T R_{M-1,i} u_{M-1,i} \notag \\
  =  & \Big ( \sum_{j \in \bar{\mathcal{N}}_i } \omega_{ji} \bar{P}_{M,i}^{-1}  P_{M,j}  A_{M-1} x_{M-1,j} \Big )^T  \bar{P}_{M,i} \Big (* \Big ) \notag \\
  \overset{\text{d}}{\leq} &   \sum_{j \in \bar{\mathcal{N}}_i } d_{i}^{out} \omega_{ji}^2 x_{M-1,j}^T A_{M-1}^T P_{M,j} \bar{P}_{M,i}^{-1}  P_{M,j} A_{M-1} x_{M-1,j}   \notag \\
  \overset{\text{e}}{\leq} &   \sum_{j \in \bar{\mathcal{N}}_i }  \omega_{ji} x_{M-1,j}^T A_{M-1}^T P_{M,j} \bar{P}_{M,i}^{-1}  P_{M,j} A_{M-1} x_{M-1,j} ,
\end{align}
where $ ``  \overset{\text{d}}{\leq}"$  is derived by using Lemma \ref{lemma1}, in which  $ \omega_{ji} \bar{P}_{M,i}^{-1}  P_{M,j}  A_{M-1} x_{M-1,j}$ and $ \bar{P}_{M,i} $ correspond to $\alpha_i$ and $P$, respectively, and $ ``  \overset{\text{e}}{\leq}"$  is derived based on the condition $ \omega_{ij} \in (0, 1/d_{j}^{out}]$, $j \in \mathcal{N}_i$ (equivalently $ \omega_{ji} \in (0, 1/d_{i}^{out}]$, $i \in \mathcal{N}_j$).

Further, $\hat{J}_{M}$ in (\ref{equ:Hat_J}) satisfies
\begin{small}
 \begin{align}
 & \hat{J}_{M} - \hat{J}_{M-1} \notag \\
  \leq  &  \sum_{i=1}^{N} \sum_{j \in \bar{\mathcal{N}}_i } x_{M-1,j}^T A_{M-1}^T (  \omega_{ji}  P_{M,j} \bar{P}_{M,i}^{-1}  P_{M,j}) A_{M-1} x_{M-1,j} \notag \\
  \overset{e}{=}  & \sum_{i=1}^{N} \sum_{j \in \mathcal{N}_i } x_{M-1,i}^T A_{M-1}^T (  \omega_{ij}  P_{M,i} \bar{P}_{M,j}^{-1}  P_{M,i}) A_{M-1} x_{M-1,i} \notag \\
  \overset{f}{=}  & \sum_{i=1}^{N}  x_{M-1,i}^T A_{M-1}^T ( P_{M,i} P_{M,i}^{-1}  P_{M,i}) A_{M-1} x_{M-1,i} \notag \\
  =  & \sum_{i=1}^{N}  x_{M-1,i}^T A_{M-1}^T  P_{M,i} A_{M-1} x_{M-1,i} \notag
\end{align}
\end{small}
where $ `` \overset{e}{=} " $ is derived based on the relation between $\mathcal{G}$ and $\bar{\mathcal{G}}$ that is revealed above  (\ref{system_model11}), and $ `` \overset{e}{=} " $ is derived from (\ref{equ:1P}). Then, by mathematical induction, one has
\begin{small}
 \begin{align}
 &  \hat{J}_{M}
  \leq     \hat{J}_{M-1}  +  \sum_{i=1}^{N}  x_{M-1,i}^T A_{M-1}^T  P_{M,i} A_{M-1} x_{M-1,i} \notag \\
  =   &  \hat{J}_{M-2}  +  \sum_{i=1}^{N} \Big ( x_{M-1,i}^T \breve{P}_{M-1,i} x_{M-1,i} + u_{M-2,i}^T R_{M-2,i} u_{M-2,i}  \Big ) \notag  \\
  \leq   &  \hat{J}_{M-3}  +  \sum_{i=1}^{N} \Big (  x_{M-2,i}^T \breve{P}_{M-2,i} x_{M-2,i} + u_{M-3,i}^T R_{M-3,i} u_{M-3,i}  \Big ) \notag  \\
  \leq   &  \cdots \notag \\
  \leq   &   \sum_{i=1}^{N}  x_{0,i}^T \breve{P}_{0,i} x_{0,i} . \notag
\end{align}
\end{small}

Notice that the initial value $ x_{0,i}$ is set as $ \frac{1}{N} x_{0} $, thus $ x_{0,i}^T \breve{P}_{0,i} x_{0,i} = {1} / {N^2} x_{0}^T \breve{P}_{0,i} x_{0}$ holds. Hence, it follows from the above inequality that
 \begin{align} \label{M_bound}
 &  \hat{J}_{M}
  \leq    \frac{1}{N^2} \sum_{i=1}^{N}   x_{0}^T \breve{P}_{0,i} x_{0}.
\end{align}
Now, the proof of Theorem \ref{thm2} is complete.

\section{Proof of Theorem \ref{thm3}}\label{proof3}
Without loss the generality, it suffices to guarantee the convergence of $\breve{P}_{k,i}$. In the following, the monotonicity of $\breve{P}_{k,i}$ is proved.

First, it follows from (\ref{equ:1overlineP})-(\ref{equ:1breveP}) that matrices $\breve{P}_{k,i}$, $\bar{P}_{k,i}$ and $P_{k,i}$ are positive definite, thus they are nonsingular. Then, if Assumptions \ref{asm:Invertibility} and \ref{asm:Time-invariant} hold, a  relation  between $(\breve{P}_{k-1,i}-NQ)^{-1}$ and $(\breve{P}_{k,i}-NQ)^{-1}$ can be revealed as
 \begin{align}
  & (\breve{P}_{k-1,i}-NQ)^{-1} - (\breve{P}_{k,i}-NQ)^{-1} \notag \\
  =  & A^{-1} ( P_{k,i}^{-1} - P_{k+1,i}^{-1}) A^{-T} \notag \\
  =  & A^{-1}  \Big \{ \sum_{j \in \mathcal{N}_i }   \omega_{ij}  (\bar{P}_{k,j}^{-1} - \bar{P}_{k+1,j}^{-1}) \Big \} A^{-T} \notag \\
  = & A^{-1} \Big \{ \sum_{j \in \mathcal{N}_i } \omega_{ij}  (\breve{P}_{k,j}^{-1} - \breve{P}_{k+1,j}^{-1}) \Big \} A^{-T}, \notag
\end{align}
which indicates that  $\breve{P}_{k-1,i}>\breve{P}_{k,i}$ if $\breve{P}_{k,i}>\breve{P}_{k+1,i}$. Besides, from (\ref{equ:1overlineP}) and (\ref{equ:1P}), $P_{M} > 0$ when $\breve{P}_{M,i} = N Q > 0$. Hence, $\breve{P}_{M-1,i}=A^{T} P_{M} A + NQ >   \breve{P}_{M,i}$. By mathematical induction, $\breve{P}_{k-1,i}>\breve{P}_{k,i}$ holds, $\forall i \in \mathcal{N} $, $\forall k = 0,1,\ldots,M$. Therefore, $\breve{P}_{k,i}$ is monotonically increasing as $k$ goes to zero from infinity.

Considering that $\breve{P}_{k,i}$ is upper bounded, it can be derived that $\breve{P}_{k,i}$ converges to the value shown in Theorem \ref{thm3}.

Thus, the proof of Theorem \ref{thm3} is complete.

\section{Proof of Theorem \ref{thm4}}\label{proof4}

To prove Theorem \ref{thm4}, it suffices to justify the case where $M$ in (\ref{M_bound}) tends to infinity.

First, from Theorem \ref{thm3}, if Assumptions \ref{asm:Invertibility} and \ref{asm:Time-invariant} hold, $\lim_{M \rightarrow \infty} \breve{P}_{0,i} = \breve{P}_{i} $.

Then, it follows from (\ref{M_bound}) that
 \begin{align}
   \lim_{M \rightarrow \infty} \hat{J}_{M}
  \leq  &  \lim_{M \rightarrow \infty} \frac{1}{N^2} \sum_{i=1}^{N}   x_{0}^T \breve{P}_{0,i} x_{0} \notag \\
   = &   \frac{1}{N^2} \sum_{i=1}^{N}   x_{0}^T   (\lim_{M \rightarrow \infty} \breve{P}_{0,i}) x_{0} \notag \\
   =  & \frac{1}{N^2} \sum_{i=1}^{N}   x_{0}^T   \breve{P}_{i} x_{0} . \notag
\end{align}

Further, by combining with (\ref{equ:Hat_Jleq}), one has
\begin{align}
 &  J_{\infty}  \leq    J^{bound}_{\infty} \triangleq   \frac{1}{N^2} \sum_{i=1}^{N} x_0^T  P_{i} x_0. \notag
\end{align}

Now, the proof of Theorem \ref{thm4} is complete.

\section{Proof of Corollary \ref{corollary1}}\label{corollary1proof}

It follows from Theorems \ref{thm3} and \ref{thm4} that
 \begin{align}
& \lim_{M \rightarrow \infty} J_M - J_{M-1} \notag \\
=  & \lim_{M \rightarrow \infty} ( x_{M}^T Q x_{M} +  u_{M-1}^T R_{M-1} u_{M-1}) \notag \\
=&  \frac{1}{N^2} \sum_{i=1}^{N} x_0^T  \breve{P}_{i} x_0 - \frac{1}{N^2} \sum_{i=1}^{N} x_0^T  \breve{P}_{i} x_0 \notag \\
= & 0. \notag
\end{align}
 Since $x_{M}^T Q x_{M}$ and $u_{M-1}^T R_{M-1} u_{M-1}$ are positive, $Q > 0$ and $R_{M-1} > 0$, the above equation indicates that $\lim_{M \rightarrow \infty} x_{M} = 0$.

\bibliographystyle{IEEEtran}
\bibliography{ref}

\end{document}